\documentstyle[12pt]{article}
\title{\bf A Solution of the Cauchy Problem for the Loop Equation in 
Turbulence}
\author{D.V.ANTONOV \thanks{E-mail: antonov@pha2.physik.hu-berlin.de; 
on leave of absence from the Institute of Theoretical and Experimental 
Physics (ITEP); supported by Graduiertenkolleg {\it Elementarteilchenphysik}, 
Russian Fundamental Research Foundation, Grant No.96-02-19184, DFG-RFFI, 
Grant 436 RUS 113/309/0 and by the INTAS, Grant No.94-2851.}
\\
{\it Institut f\"ur Elementarteilchenphysik, Humboldt-Universit\"at zu 
Berlin,}\\
{\it Invalidenstrasse 110, D-10115, Berlin, Germany}\\}
\date{}
\begin{document}
\maketitle
\vspace{1cm}
\centerline{\bf {Abstract}}

\vspace{3mm}
Under certain conditions, imposed on the viscosity of the fluid, initial 
data and the class of contours under consideration, the Cauchy problem 
with finite values of time for 
the loop equation in turbulence with Gaussian random forces is solved by 
making use of the smearing procedure for the loop space functional 
Laplacian. The solution obtained depends on the initial data and its 
functional derivatives and on the potential of the random forces.

\vspace{1cm}

Nowadays there exist various field theoretical approaches to the problem 
of turbulence. The main one, which exploits the methods of conformal field 
theory$^{1}$, was proposed in Ref. 2 and developed in Ref. 3. In Ref. 4 
the operator 
product expansion method was applied to investigation of the so-called 
Burgers' turbulence, i.e. one-dimensional turbulence without pressure 
(see also Ref. 5, where alternative approaches to this problem such as the 
instanton approach and replica method were suggested). Field 
theoretical methods were also used in Ref. 6 in order to calculate the 
probability distribution and develop Feynman diagrammatic technique 
in the theory of wave turbulence.

An approach to the problem of turbulence, based on the loop 
calculus$^{7,8}$ 
(for a review see Ref. 9) was suggested in Ref. 10 and developed in Ref. 11 
(see also Ref. 12, where its relation to the generalized Hamiltonian 
dynamics and the Gibbs-Boltzmann statistics was established). Within 
this approach one deals with loop functionals of the Stokes type 
$\Psi(C)=\left<exp\left(\frac{i}{\nu}\oint\limits_C^{}d r_
\alpha v_\alpha\right)\right>_
{\vec f}$, where $\vec v$ is the velocity of the fluid, $\nu$ is its 
viscosity, and $\vec f$~'s are external Gaussian random 
forces, whose 
bilocal correlator reads

$$\left<f_\alpha \left(\vec r, t\right) f_\beta\left(\vec r{\,}^\prime
, t^\prime\right)\right>_{\vec f}=
\delta_{\alpha\beta}\delta(t-t^\prime)F\left(\left(\vec r-\vec r{\,}^
\prime\right)^2\right).$$
Taking into account that $\vec v$ satisfies the Navier-Stokes equations

$$\dot{\vec v}=\nu\nabla^2\vec v -\left(\vec v{\,}\nabla\right)\vec v -
\nabla p-\vec f,~~\nabla\vec v=0,$$
where $p$ is the pressure of the fluid, 
one can write down the following equation$^{11}$ 
for the functional $\Psi(C)$

$$i\nu\dot\Psi(C)=\left(\nu^2\oint\limits_C^{}d r_\alpha\left(i\partial_\beta^
{\vec r}\frac{\delta}{\delta\sigma_{\beta\alpha}(\vec r{\,})}+\int
d^3r^\prime\frac{r_\gamma^\prime-r_\gamma}{4\pi\left|\vec r-\vec r{\,}^
\prime\right|^3}
\frac{\delta^2}{\delta\sigma_{\beta\alpha}\left(\vec r{\,}\right)
\delta\sigma_{\beta
\gamma}\left(\vec r{\,}^\prime\right)}\right)+iU(C)\right)\Psi(C), 
\eqno (1)$$
where $\partial_\beta^{\vec r(\sigma)}\equiv\int\limits_{\sigma-0}^
{\sigma+0}d \sigma^\prime\frac{\delta}{\delta r_\beta (\sigma^\prime)}, 
iU(C)\equiv\frac{i}{\nu}\oint\limits_C^{}d r_\alpha\oint\limits_C^{} d 
r_\alpha^\prime F\left(\left(\vec r-\vec r{\,}^\prime\right)^2\right)$ 
is the imaginary
potential term, and 
the gradient of the 
pressure does not contribute to the operator standing on the R.H.S. of 
Eq. (1). Since the functional $\Psi(C)$ does not have marked points, Eq. (1) 
may be rewritten in the following way

$$\left(i\nu\frac{\partial}{\partial t}-i\nu^2\Delta-iU(C)\right)
\Psi(C)=\nu^2\oint\limits_C^{}dr_\alpha\int d^3r^\prime\frac{r_\gamma^
\prime-r_\gamma}
{4\pi\left|\vec r-\vec r{\,}^\prime\right|^3}\frac{\delta^2\Psi(C)}{\delta
\sigma_{\beta\alpha}\left(\vec r{\,}\right)\delta\sigma_{\beta\gamma}
\left(\vec r{\,}^\prime\right)}, \eqno (2)$$
where $\Delta\equiv\int\limits_0^1 d\sigma\int\limits_{\sigma-0}^
{\sigma+0} d\sigma'\frac{\delta^2}{\delta r_\mu\left(\sigma^\prime\right)
\delta r_\mu\left(\sigma\right)}$ is the functional Laplacian.
     
In this letter we shall solve the Cauchy problem with finite 
values of time for Eq. (2) in the case of some special initial condition 
and certain restrictions, imposing on the viscosity of the fluid and on 
the class of contours $C$ under consideration. Namely, let us look for a 
solution, annihilating the R.H.S. of Eq. (2). This assumption will 
finally yield the constraints, mentioned above. 

In order to invert the operator standing on the L.H.S. of Eq. (2), we   
shall make use of the method suggested 
in Ref. 13, where it was shown that for an arbitrary functional $Q\left[
r\right]$, defined on the loop space, the following 
equation holds true 

$$\Delta^{(G)}\left<Q\left[r+\sqrt{A}\xi\right]\right>
_{\xi}^{(G)}=
2\frac{d}{dA}\left<Q\left[r+\sqrt{A}\xi\right]\right>
_{\xi}^{(G)}, \eqno (3)$$
where $r\equiv\vec r(s),~0\le s\le 1,$ stands for an element 
of the loop space, the smeared functional Laplacian has the form

$$\Delta^{(G)}=\int\limits_0^1 d\sigma~v.p.\int\limits_0^1 d\sigma^\prime 
G\left(\sigma-\sigma^\prime\right)\frac{\delta^2}{\delta r_\mu\left(
\sigma^\prime\right)
\delta r_\mu\left(\sigma\right)}+\Delta, \eqno (4)$$
and the average over loops is defined as follows 

$$\left<Q\left[\xi\right]\right>_{\xi}^{(G)}=
\frac{\int\limits_{\vec\xi(0)=
\vec\xi(1)}^{}D\xi e^{-S}Q\left[\xi\right]}
{\int\limits_{\vec
\xi(0)=\vec\xi(1)}^{}D\xi e^{-S}}. \eqno (5)$$
Here $G\left(\sigma-\sigma^\prime\right)$ is a certain smearing function,

$$S=\frac{1}{2}\int\limits_0^1 d\sigma\int\limits_0^1 d\sigma^
\prime\vec\xi\left(\sigma\right)G^{-1}\left(\sigma-\sigma^\prime
\right)\vec\xi\left(\sigma^
\prime\right), \eqno (6)$$
and the first term on the R.H.S. of Eq. (4) is an operator of the second 
order (does not satisfy the Leibnitz rule in contrast to the operator 
$\Delta$) and is reparametrization noninvariant. 
In particular, for 

$$G\left(\sigma-\sigma^\prime\right)=e^{-\frac{\left|\sigma-
\sigma^\prime\right|}
{\varepsilon}},~
\varepsilon\ll 1 \eqno (7)$$
action (6) becomes local: $S=\frac{1}{4}\int\limits_0^1 
d\sigma\left(\varepsilon\dot{{\vec\xi}}{\:}^2(\sigma)+\frac{1}{\varepsilon}
{\vec\xi}{\:}^2(\sigma)\right)$. This is the action of the Euclidean harmonic 
oscillator at finite temperature. For $G\left
(\sigma-\sigma^\prime\right)$ defined
by Eq. (7) the contribution of the first term on the R.H.S. of Eq. (4) is of 
order $\varepsilon$ for smooth contours, and therefore 
when $\varepsilon\to 0$ it vanishes,   
$\Delta^{(G)}$ tends to the functional Laplacian 
$\Delta$, and reparametrization invariance restores. Eq. (3) may be 
easily obtained from the equation of motion

$$\left<\xi_\mu(\sigma)Q\left[\xi\right]\right>
_{\xi}^{(G)}=
\int\limits_0^1 d\sigma^\prime G\left(\sigma-\sigma^\prime\right)\left<\frac{
\delta Q\left[\xi\right]}{\delta\xi^\mu\left(\sigma^\prime
\right)}\right>_
\xi^{(G)}, $$
which follows from Eqs. (5) and (6). 

Making a shift of the contour $C, \vec r\to\vec r+\sqrt{A}\vec \xi,$ 
in Eq. (2) with the vanishing R.H.S. and  
averaging this equation over loops according to formulae 
(5) and (6), we arrive by virtue of Eq. (3) with $Q=\Psi$ at the following 
solution of the Cauchy problem for this equation 

$$\left<\Psi\left[r+\sqrt{A}\xi, t\right]\right>_
\xi=\left<\Psi\left[r+\sqrt{A+2\nu t}\xi, 0\right]\exp\left(\frac{1}{\nu}
\int\limits_0^t d\tau U\left[r+\sqrt{A+2\nu\tau}\xi\right]\right)\right>
_\xi. \eqno (8)$$
Notice that in the case when $t$ may take arbitrary values, in order that 
this solution to be bounded at $t$ tending to infinity, one should demand 
that the potential $U[r]$ satisfy the following condition: for an arbitrary 
nonnegative number $a$ and for an arbitrary element $\rho$ of the loop space 
the following inequality should hold $\int\limits_0^a d\lambda U[\lambda
\rho]\le 0$.

Let us now turn ourselves to the case of finite values of time and eliminate 
the $\xi$-averaging in Eq. (8). To this end we shall make use of the 
following formula$^{14}$

$$\left<e^B D\right>=\left(\exp\left(\sum\limits_{n=1}^{\infty}\frac{1}
{n!}\left<\left< B^n\right>\right>\right)\right)\left(\left<D\right>+\sum
\limits_{k=1}^\infty \frac{1}{k!}\left<\left<B^k D\right>\right>\right), 
\eqno (9)$$
where $B$ and $D$ are two statistically dependent commuting quantities, and 
$\left<\left<...\right>\right>$ denotes the so-called cumulants, i.e. 
irreducible correlators. Applying formula (9) to Eq. (8), we see that one 
can neglect all the cumulants higher than quadratic in functionals 
$\Psi$ and $U$ (this is the so-called bilocal approximation$^{15}$) if 
the following inequality holds

$$\alpha\equiv\frac{tS_{min.}max\left(\left|F\right|\right)}
{\nu^2}\ll 1, \eqno (10)$$
where $S_{min.}$ is the area of the minimal surface encircled by the 
contour $C$ ($\sqrt{S_{min.}}$ is a measure of the loop size$^{11}$). Let 
us also assume that there exists another small parameter in the problem 
under consideration,

$$\beta\equiv\frac{t\nu}{S_{min.}}\ll 1, \eqno (11)$$
so that one can, for example, approximately write down the following 
formula

$$\left<\Psi\left[r+\sqrt{A+2\nu t}\xi, 0\right]\right>_\xi=\left<
\Psi\left[r+\sqrt{A}\xi, 0\right]\right>_\xi+\nu t\Delta^{(G)}
\left<\Psi\left[r+\sqrt{A}\xi, 0\right]\right>_\xi,$$   
where all the terms on the R.H.S. higher than those of the first order 
in $\beta$ are neglected. Notice that condition (11) is introduced only 
in order that one would be able not to take into account these terms, 
which contain unfamiliar operators higher than of the second order 
in functional derivatives. Otherwise such terms will appear in 
final solution (12).

Then, taking into account that $\Delta^{(G)}U[r]=0$, expanding the 
R.H.S. of Eq. (8) by virtue of formula (9) up to the terms not higher 
than of the second order in $\alpha$ and of the first order in $\beta$ 
and putting finally $A$ being equal to zero, we arrive at the following 
solution of the Cauchy problem for Eq. (2)  

$$\Psi[r,t]=\Biggl(\left(1+\frac{t}{\nu}U[r]+\frac{t^2}{2\nu^2}U^2[r]
\right)\left(1+t\nu\Delta^{(G)}\right)+$$

$$+\frac{t^3}{2\nu}\int\limits_0^1
d\sigma\int\limits_0^1 d\sigma'G\left(\sigma-\sigma'\right)\left(
\frac{\delta}{\delta r_\alpha\left(\sigma\right)}U[r]\right)
\left(\frac{\delta}
{\delta r_\alpha\left(\sigma'\right)}U[r]\right)\Biggr)\Psi[r,0]+$$

$$+\frac{3t^2}{2}\left(1+\frac{t}{\nu}U[r]\right)\int\limits_0^1 d\sigma
\int\limits_0^1 d\sigma'G\left(\sigma-\sigma'\right)\left(\frac{\delta}
{\delta r_\alpha\left(\sigma\right)}U[r]\right)\left(\frac{\delta}
{\delta r_\alpha\left(\sigma'\right)}\Psi[r,0]\right),$$
which at $\varepsilon$ tending to zero takes the form

$$\Psi[r,t]=\left(1+\frac{t}{\nu}U[r]+\frac{t^2}{2\nu^2}U^2[r]\right)
\left(1+t\nu\Delta\right)\Psi[r,0]. \eqno (12)$$

Now we shall derive the constraints, imposing on the initial condition 
and on the class of contours $C$ under consideration in order that the 
obtained solution (12) would really annihilate the R.H.S. of Eq. (2). 
Taking into account that the area derivative satisfies the Leibnitz 
rule$^{9}$, and that$^{8}$ $\frac{\delta U(C)}{\delta\sigma_{\mu\nu}
\left(\vec r\left(s\right)\right)}=\frac{1}{\nu}\dot t_{[\mu}(s)t_{\nu]}(s)
\int\limits_{-\infty}^{+\infty}dzF\left(z^2\right)$, where $t_\mu$ is the 
local tangent vector of the contour, i.e. $dr_\mu(s)=t_\mu(s)ds$, and the 
square brackets stand for the antisymmetrization, we arrive at the 
following constraints

$$\oint\limits_C^{}dr_\alpha'\int d^3r''\frac{r_\gamma'-r_\gamma''}
{\left|\vec r{\,}^\prime-\vec r{\,}^{\prime\prime}
\right|^3}\frac{\delta^2\Psi[r,0]}{\delta\sigma
_{\beta\alpha}\left(\vec r{\,}^\prime\right)\delta\sigma_{\beta\gamma}
\left(\vec r{\,}^{\prime\prime}\right)}=0, $$

$$\oint\limits_C^{}dr_\alpha\int d^3r'\frac{r_\gamma-r_\gamma'}{\left|
\vec r-\vec r{\,}^\prime\right|^3}\left(\dot t_{[\beta}\left(\vec r{\,}
\right)t_{\alpha]}\left(\vec r{\,}\right)\right)\left(\dot t_{[\beta}
\left(\vec r{\,}^\prime\right)t_{\gamma]}\left(\vec r{\,}^\prime\right)
\right)=0 \eqno (13)$$   
and 

$$\oint\limits_C^{}dr_\alpha'\int d^3r''\frac{r_\gamma'-r_\gamma''}
{\left|\vec r{\,}^\prime-\vec r{\,}^{\prime\prime}\right|^3}\left(
\dot t_{[\beta}\left(\vec r{\,}^\prime\right)t_{\alpha]}\left(\vec r
{\,}^\prime\right)\frac{\delta\Psi[r,0]}{\delta\sigma_{\beta\gamma}
\left(\vec r{\,}^{\prime\prime}\right)}+\dot t_{[\beta}\left(\vec r{\,}^
{\prime\prime}\right)t_{\gamma]}\left(\vec r{\,}^{\prime\prime}\right)
\frac{\delta\Psi[r,0]}{\delta\sigma_{\beta\alpha}\left(\vec r{\,}^\prime
\right)}\right)=0. $$
Here, of course, $t_\alpha\left(\vec r{\,}^\prime\right)=
\dot t_\alpha\left(\vec r{\,}^\prime\right)=\frac{\delta\Psi[r,0]}
{\delta\sigma_{\alpha\beta}\left(\vec r{\,}^\prime\right)}=0$ 
if $\vec r{\,}^\prime$ does not belong 
to the contour $C$. When being accounted for, higher cumulants will bring 
into solution (12) terms with the space dependence of the form $U^n[r]
\Psi[r,0]$ and $U^n[r]\Delta\Psi[r,0]$, where $n$ is an integer. Since the 
Leibnitz rule is true for the area derivative, it is easy to show that 
all these terms will also annihilate the R.H.S. of Eq. (2) if constraints 
(13) hold, so that no any other additional constraints are needed.

Notice, that if the initial condition satisfies the first of constraints 
(13) in a ``strong'' sense, i.e. if $\frac{\delta^2\Psi[r,0]}{\delta\sigma_
{\beta\alpha}\left(\vec r{\,}^\prime\right)\delta\sigma_{\beta\gamma}
\left(\vec r{\,}^{\prime\prime}\right)}=0$, which takes place, for example, 
when 

$$\Psi[r,0]=Q\left[\oint\limits_C^{}dr_\alpha'\oint\limits_C^{}
dr_\alpha''\Phi\left(\left(\vec r{\,}^\prime-\vec r{\,}^{\prime\prime}
\right)^2\right)\right], \eqno (14)$$
where $Q$ is an arbitrary functional defined on the loop space, and $\Phi$ 
is an arbitrary function, which provides the convergence of the double 
integral on the R.H.S. of Eq. (14), then $\Delta\Psi[r,0]=0$, since, as it 
was shown in Ref. 16, $\Delta=\oint\limits_C^{}dr_\alpha'\oint\limits_
{C_{\vec r{\,}^\prime\vec r{\,}^\prime}}^{}dr_\gamma''\frac{\delta^2}
{\delta\sigma_{\beta\alpha}\left(\vec r{\,}^\prime\right)\delta\sigma_
{\gamma\beta}\left(\vec r{\,}^{\prime\prime}\right)}$. This means that 
according to Eq. (12) $\Delta\Psi[r,t]$ also vanishes, and therefore 
$\Psi[r,t]$ should be a solution of the Cauchy problem for degenerated 
Eq. (2) with the vanishing R.H.S., which has the form $\left(\nu\frac{
\partial}{\partial t}-U[r]\right)\Psi[r,t]=0$. This solution obviously 
reads as $\Psi[r,t]=\Psi[r,0]e^{\frac{t}{\nu}U[r]}$, and hence it 
annihilates the R.H.S. of Eq. (2) by virtue of constraints (13) 
(as it is explained in the previous paragraph) and actually satisfies 
the condition $\Delta\Psi[r,t]=0$.   

In conclusion, we have solved the Cauchy problem with finite values 
of time for the loop equation (2). When inequalities (10) and (11) and 
constraints (13) hold, the solution is given by formula (12) up to the terms 
linear in the parameter $\beta$ (which corresponds to accounting 
in solution (12) only 
for the terms, which do not contain functional derivatives higher than 
of the second order) and quadratic in $\alpha$ (which corresponds to the 
bilocal approximation in the cumulant expansion of the R.H.S. of Eq. (8)).
Among constraints (13) the first one is imposed on the initial condition, 
the second one is imposed on the class of contours $C$ under 
consideration, and the last one is imposed on both of them. This set of 
constraints is enough in order that all possible terms, bringing into 
solution (12) by higher cumulants, would also annihilate the R.H.S. of 
Eq. (2).

\vspace{6mm}
{\large \bf Acknowledgments}

\vspace{3mm}
I am grateful to Professors Yu.M.Makeenko and Yu.A.Simonov for the 
useful discussion and to Professor Yu.M.Makeenko for bringing my 
attention to 
Ref. 13. I would also like to thank the theory group of the Quantum 
Field Theory Department of the Institut f\"ur Physik of the 
Humboldt-Universit\"at of Berlin for kind hospitality.

\newpage
{\large\bf References}

\vspace{3mm}
\noindent
1.~ A.A.Belavin, A.M.Polyakov and A.B.Zamolodchikov, 
{\it Nucl.Phys.} {\bf B241}, 333 (1984).\\
2.~ A.M.Polyakov, {\it Nucl.Phys.} {\bf B396}, 367 (1993).\\
3.~ G.Ferretti and Z.Yang, {\it Europhys.Lett.} {\bf 22}, 639 (1993); 
G.Falkovich and A.Hanany, Preprint WIS-92-88-PH, {\it hep-th}/9212015 (1992); 
H.Cateau, Y.Matsuo and M.Umeki, Preprint UT-652, {\it hep-th}/9310056 (1993); 
B.K.Chung, Soonkeon Nam, Q-Han Park and H.J.Shin, 
{\it Phys.Lett.} {\bf B309}, 58 (1993), {\it Phys.Lett.} 
{\bf B317}, 92 (1993), {\it Phys.Lett.} {\bf B317}, 97 (1993); 
G.Falkovich and A.Hanany, {\it Phys.Rev.Lett.} {\bf 71}, 3454 (1993); 
Y.Matsuo, {\it Mod.Phys.Lett.} {\bf A8}, 619 (1993); 
D.Lowe, {\it Mod.Phys.Lett.} {\bf A8}, 923 (1993); 
C.Chen, {\it Mod.Phys.Lett.} {\bf A9}, 123 (1994); 
Ph.Brax, Preprint SPhT 95/058 DAMTP 95/25, {\it hep-th}/9505111 (1995); 
L.Moriconi, Preprint PUPT-1554, {\it hep-th}/9508040 (1995); 
M.R.Rahimi Tabar and S.Rouhani, {\it hep-th}/9507166; 
O.Coceal and S.Thomas, {\it Mod.Phys.Lett.} {\bf A10}, 2427 (1995), 
{\it Int.J.Mod.Phys.} {\bf A11}, 5261 (1996); 
O.Coceal, Wafic A.Sabra and S.Thomas, 
Preprint QMW-PH-96-05, {\it hep-th}/9604157 (1996); 
M.R.Rahimi Tabar and S.Rouhani, {\it Ann.Phys.} {\bf 246}, 446 (1996), 
Preprint IPM-96-150, {\it hep-th}/9606143 (1996), Preprint IPM-96A, 
{\it hep-th}/9606154 (1996); 
Ph.Brax, Preprint DAMTP/96-38 SPhT/96-033, {\it hep-th}/9606156 (1996), 
Preprint DAMTP/96-56 SPhT/96-60, {\it hep-th}/9607230 (1996); 
Michael A.I.Flohr, Preprint IASSNS-HEP-96/69, {\it hep-th}/9606130 
(1996).\\
4.~ A.M.Polyakov, {\it Phys.Rev.} {\bf E52}, 6183 (1995); 
E.Elizalde, Preprint UTF355, {\it hep-th}/9508097 (1995); 
M.R.Rahimi Tabar, S.Rouhani and B.Davoudi, {\it hep-th}/9507165; 
S.Boldyrev, {\it hep-th}/9610080.\\
5.~ J.-P.Bouchaud et al., {\it Phys.Rev.} {\bf E52}, 3656 (1995); 
G.Falkovich, I.Kolokolov, V.Lebedev and A.Migdal, {\it chao-dyn}/9512006; 
V.Gurarie and A.A.Migdal, {\it hep-th}/9512128; 
E.Balkovsky, G.Falkovich, I.Kolokolov and V.Lebedev, {\it chao-dyn}/9603015; 
J.-P.Bouchaud and M.M\'ezard, Preprint LPTENS 96/37, {\it cond-mat}/
9607006 (1996).\\
6.~ V.Gurarie, Preprint PUPT-1525, {\it hep-th}/9501021 (1995), 
{\it Nucl.Phys.} {\bf B441 [FS]}, 569 (1995).\\
7.~ A.M.Polyakov, {\it Nucl.Phys.} {\bf B164}, 171 (1980).\\
8.~ Yu.M.Makeenko and A.A.Migdal, {\it Nucl.Phys.} {\bf B188}, 269 (1981).\\
9.~ A.A.Migdal, {\it Phys.Rep.} {\bf 102}, 199 (1983).\\ 
10. A.A.Migdal, Preprint PUPT-1383, {\it hep-th}/9303130 (1993).\\
11. A.A.Migdal, {\it Int.J.Mod.Phys.} {\bf A9}, 1197 (1994).\\
12. A.A.Migdal, Preprint PUPT-1409, {\it hep-th}/9306152 (1993).\\
13. Yu.M.Makeenko, {\it Phys.Lett.} {\bf B212}, 221 (1988), preprints 
ITEP 88-50, ITEP 89-18; unpublished.\\ 
14. N.G. Van Kampen, {\it Stochastic Processes in Physics and Chemistry}
(North-Holland Physics Publishing, 1984).\\
15. H.G.Dosch, {\it Phys.Lett.} {\bf B190}, 177 (1987); Yu.A.Simonov, 
{\it Nucl.Phys.} {\bf B307}, 512 (1988); H.G.Dosch and Yu.A.Simonov, 
{\it Phys.Lett.} {\bf B205}, 339 (1988), {\it Z.Phys.} {\bf C45}, 147 
(1989); Yu.A.Simonov, {\it Nucl.Phys.} {\bf B324}, 67 (1989), {\it 
Phys.Lett.} {\bf B226}, 151 (1989), {\it Phys.Lett.} {\bf B228}, 413 
(1989), {\it Yad.Fiz.} {\bf 54}, 192 (1991); 
D.V.Antonov and Yu.A.Simonov, {\it Int.J.Mod.Phys.} {\bf A11}, 4401 
(1996); D.V.Antonov, D.Ebert and Yu.A.Simonov, {\it Mod.Phys.Lett.} 
{\bf A11}, 1905 (1996); D.V.Antonov, {\it JETP Lett.} {\bf 63}, 398 (1996), 
{\it Int.J.Mod.Phys.} {\bf A12}, 2047 (1997), {\it Yad.Fiz.} {\bf 60}, 
365, 553 (1997).\\
16. A.A.Migdal, {\it Nucl.Phys.} {\bf B189}, 253 (1981). 
\end{document}